\documentclass{llncs} 

\newtheorem{defn}{D\'efinition}[section]

\newtheorem{thm}[defn]{Theorem}
\newtheorem{lem}[defn]{Lemma}
\newtheorem{cor}[defn]{Corollary}

\begin{document}

\title{A POLYNOMIAL TIME ALGORITHM FOR  LEFT [RIGHT] LOCAL TESTABILITY}
\author{A.N. Trahtman}
\date{}
\institute{Bar-Ilan University, Dep. of Math. and CS, 52900,Ramat
Gan, Israel  email:trakht@macs.biu.ac.il}
 \maketitle

\centerline{Lecture Notes in Computer Scence 2608(2003), 203-212}

\begin{abstract} A right [left] locally testable language $S$ is a language with
        the property that for some nonnegative integer  $k$
        two words $u$ and $v$ in alphabet $S$
    are equal in the semigroup if  (1) the prefix and
        suffix of the words of length $k-1$ coincide, (2) the  set of
        segments of length $k$ of the words
 as well as 3) the order of the first appearance of these segments in prefixes
 [suffixes] coincide.

We present necessary and sufficient condition for graph [semigroup] to be
transition
 graph [semigroup] of the deterministic finite automaton that accepts
right [left] locally testable language and necessary and sufficient
condition for transition graph of the deterministic finite automaton
with locally idempotent semigroup.
      We introduced polynomial time algorithms for the right
 [left] local testability problem for transition semigroup and transition
 graph of the deterministic finite automaton based on these conditions.
Polynomial time algorithm verifies transition graph of automaton
with locally idempotent transition semigroup.
\end{abstract}



 \section{Introduction}

   The concept of local testability was first introduced by McNaughton
 and Papert { \cite {MP}} and by Brzozowski and Simon { \cite
 {BS}}. This concept is
 connected with languages, finite automata and semigroups and has
a wide spectrum of generalizations.

  The  necessary and sufficient condition for local testability
were investigated for both transition graph  and transition
 semigroups of the automaton {\cite {BS}, \cite {K91}, \cite {Ta}, \cite {Z}}.
The  polynomial time algorithms solve the problem of local
testability for transition graph {\cite {K91}} and for transition
semigroups of the automaton {\cite {Ta}}.
 They are polynomial in terms of the size of the semigroup or
in the sum of nodes and edges.

Right [left] local testability was introduced and studied by
K\"{o}nig \cite{Ko} and by Garcia and Ruiz \cite {GR}. These
papers use different definitions of the conception and we follow
\cite{GR} here
\begin{thm} \cite {GR}
 A finite semigroup $S$ is right [left] locally testable iff it is locally
idempotent and locally satisfies the identity $xyx=xy$ [$xyx=yx$].
\end{thm}
 For conception of local idempotency see, for instance, {\cite {Co}}.
  The varieties of semigroups defined by considered identities are
  located not far from atoms in the structure of idempotent
varieties { \cite {Br}}.

We present in this work necessary and sufficient condition
 for right [left] local testability for transition graph of the DFA 
 and for the local idempotency of the transition semigroup on the
corresponding transition graph.
We improve  necessary and sufficient condition
 for right [left] local testability from \cite {GR} for transition
 semigroup. On the base of these results, we
 introduced  a polynomial time algorithm for the right [left]
local testability problem for transition semigroup and transition
 graph of the deterministic finite automaton and for checking the 
transition graph of the automaton with locally idempotent semigroup.

These algorithms are implemented in the package
TESTAS. The package checks also whether or not a language given by its
minimal automaton or by syntactic semigroup of the automaton is
locally testable, threshold locally testable, strictly locally
 testable, or piecewise testable {\cite {TP}}, {\cite {TL}}.

 \section{Notation and definitions}

 Let $\Sigma$ be an alphabet and let $\Sigma^+$ denote  the free semigroup
on $\Sigma$. If $w \in \Sigma^+$, let $|w|$ denote the length of $w$.
Let $k$ be a positive integer. Let $i_k(w)$ $[t_k(w)]$ denote the prefix
[suffix] of $w$ of length $k$ or $w$ if $|w| < k$. Let $F_k(w)$ denote  the
 set of segments of $w$ of length $k$.
 A language $L$ [a semigroup $S$] is called {\it right [left] k-testable} if
there is an alphabet $\Sigma$
[and a surjective morphism $\phi : \Sigma^+ \to S$] such that  for all
$u$, $v \in \Sigma^+$, if $i_{k-1}(u)=i_{k-1}(v), t_{k-1}(u)=t_{k-1}(v)$,
 $F_k(u)=F_k(v)$ and the order of appearance of these segments in
prefixes [suffixes] in the word coincide, then either both $u$ and $v$ are in
$L$
 or neither is in $L$ [$u\phi = v\phi$].

   An automaton is  {\it right [left] $k$-testable} if the automaton
accepts a
  right [left] $k$-testable language.

   A language $L$   [a semigroup $S$, an automaton $\bf A$] is {\it right [left]
locally
 testable} if it is  right [left] $k$-testable for some $k$.

$|S|$ is the number of elements of the set $S$.

A semigroup $S$ is called {\em semigroup of left [right] zeroes}
if $S$ satisfies the identity $xy=x$ [$xy=y$].

A semigroup $S$ has a property $\rho$ {\em locally} if for any idempotent $e
\in S$
the  subsemigroup $eSe$ has the property $\rho$.

So a semigroup $S$ is called {\em locally idempotent} if
$eSe$ is an idempotent subsemigroup for any idempotent $e \in S$.

\medskip
A maximal strongly connected component of the graph will be
denoted for brevity as $\it SCC$,
a finite deterministic automaton will be denoted as $\it DFA$.
A node from an $SCC$ will be called for brevity as an $\it
SCC-node$. $SCC$-node can be defined as a node that has
a right unit in transition semigroup of the automaton.

$|\Gamma|$ denotes the number of nodes of the graph $\Gamma$.

$\Gamma^i$ denotes the direct product of $i$ copies of the
graph $\Gamma$.
The edge $({\bf p}_1,...,{\bf p}_n) \to ({\bf q}_1,...,{\bf q}_n)$
 in $\Gamma^i$ is labelled by $\sigma$ iff for each $i$ the edge
${\bf p}_i \to {\bf q}_i$ in $\Gamma$ is labelled by $\sigma$.

The graph with only trivial SCC (loops) will be called {\it acyclic}.

If an edge ${\bf p} \to \bf q$ is labelled by $\sigma$ then let us
denote the node $\bf q$ as ${\bf p}\sigma$.

 We shall write $\bf p \succeq \bf q$ if the node $\bf q$  is
reachable from the node $\bf p$ or $\bf p=q$ ($\bf p \succ \bf q$
for distinct $\bf p, q$).

In the case $\bf p \succeq q$ and $\bf q \succeq p$ we write
$\bf
p \sim q$ (that is $\bf p$ and $\bf q$ belong to one $SCC$ or $\bf
p=q$).

\section{Transition graph of deterministic finite automaton}
\subsection{Graph of DFA with locally idempotent transition semigroup}
\begin{lem} $\label {3.2}$
 Let $S$ be the transition semigroup of a
deterministic finite automaton and let $\Gamma$ be its transition graph.
Let us suppose that for three distinct nodes $\bf  p, q, r$ from $\Gamma$
the node ($\bf  p, q, r$) in  $\Gamma^3$ is $SCC$-node,
and (${\bf p, q}) \succ (\bf q, r$) in $\Gamma^2$.

 Then  $S$ is not locally idempotent.

\end{lem}

Proof. Let us  suppose that for the nodes  $\bf  p, q, r$
 from $\Gamma$ the conditions of lemma hold.
Therefore the nodes $\bf p, q, r$ have a right unit $e=e^2$,
whence ${\bf p}e=\bf p$, ${\bf q}e=\bf q$, ${\bf r}e=\bf r$. In
view (${\bf p, q}) \succ (\bf  q, r$), there exists an element $s
\in S$ such that ${\bf p}s = \bf q$ and ${\bf q}s = \bf r$.
Therefore ${\bf p}ese = \bf q$ and ${\bf q}ese = \bf r$, whence
${\bf p}(ese)^2 = {\bf r} \ne {\bf q}= {\bf p}ese$. So  ${\bf
p}(ese)^2 \ne {\bf p}ese$ and  $(ese)^2 \ne ese$. Semigroup $eSe$
is not  an idempotent semigroup and therefore $S$ is not locally
idempotent.

\begin{lem} $\label {3.3}$
 Let $S$ be the locally idempotent transition semigroup of a
deterministic finite automaton and let $\Gamma$ be its transition graph.

 For any $SCC$-node ($\bf p, q$) $\in \Gamma^2$ and $s \in S$
 from  ${\bf p}s \succeq \bf q$ follows ${\bf q}s \succeq \bf q$.
\end{lem}

Proof. Let us consider $SCC$-node ($\bf p, q$) from $\Gamma^2$
 such that ${\bf p}s \succeq \bf q$.
The node ($\bf p, q$) has a right unit $e=e^2$,
 so ${\bf p}e=\bf p$, ${\bf q}e=\bf q$.
 For some $b \in S$ we have  ${\bf p}sb = \bf q$. We can assume $s=es$,
$b=be$. $esbe=(esbe)^2$ in locally idempotent semigroup $S$.
 Therefore ${\bf q} ={\bf p}esbe = {\bf p}(esbe)^2 ={\bf q}esbe = {\bf q}sbe$.
  Thus we have ${\bf q}s \succeq {\bf q}$.

Lemma implies
\begin{cor} $\label {3.4}$
 Let $S$ be the locally idempotent transition semigroup of a
deterministic finite automaton and let $\Gamma$ be its transition
graph.

 Let us suppose that in $\Gamma^2$ we have (${\bf p, q})
\succ (\bf q, r$) and the node ($\bf p, q$)
 is an $SCC$-node.
 Then ${\bf r} \sim \bf q$.
\end{cor}

\begin{lem} $\label {3.6}$
 Let $S$ be transition  semigroup of a deterministic
finite automaton and suppose that in $\Gamma^2$
 we have $({\bf p, q}) \succ (\bf q, p)$
for two distinct nodes $\bf p, q$.

  Then  $S$ is not locally idempotent.
\end{lem}

Proof.
We have ${\bf p}s=\bf q$ and ${\bf q}s=\bf p$
 for some $s \in S$.
So  ${\bf p}s^2={\bf p} \ne{\bf p}s=\bf q$ and
 ${\bf p}={\bf p}s^{2n} \ne{\bf p}s^{2n-1}=\bf q$.
Therefore $s^{2n} \ne s^{2n-1}$ for any integer $n$
 because of  ${\bf  p} \ne \bf q$.
Finite semigroup $S$ contains therefore non-trivial
 subgroup, whence $S$ is not locally idempotent.

\medskip
Let us formulate the necessary and sufficient conditions for graph
to be transition graph of DFA with
 locally idempotent transition semigroup.
\begin{thm} $\label {3.8}$
  Transition semigroup $S$ of a
deterministic finite automaton is locally idempotent iff

1. $({\bf p,q}) \not\succ (\bf q, p$)  in $\Gamma^2$ for any two
distinct nodes $\bf p, q$,

2. for any $SCC$-node ($\bf p, q$) $\in \Gamma^2$ and $s \in S$
 from  ${\bf p}s \succeq \bf q$ follows ${\bf q}s \succeq \bf q$ and

3. for any $SCC$-node ($\bf  p, q, r$) of  $\Gamma^3$ with
distinct components holds
  $({\bf p, q}) \not\succ (\bf q, r$)  in $\Gamma^2$.
\end{thm}

Proof. If $S$ is locally idempotent then the condition 1 follows
from lemma \ref {3.6}, condition 2 follows from lemma \ref {3.3},
condition 3 follows from lemma \ref {3.2}.

 Suppose now that $S$ is not locally idempotent.
Then for some node $\bf p$ from $\Gamma$,
 idempotent $e$ and element $s$ from $S$
we have ${\bf p}(ese)^2  \ne  {\bf p}ese$. Hence ${\bf p}e  \ne
{\bf p}ese$ and at least one of two nodes ${\bf p}(ese)^2$, ${\bf
p}ese$ exists. If exists the node ${\bf p}(ese)^2$  then the node
${\bf p}ese$ exists too. So ${\bf p}ese$ exists anyway. Therefore
${\bf p}e$ exists too and from $({\bf p}e, {\bf p}ese)ese = ({\bf
p}ese, {\bf p}(ese)^2)$ in view of condition 2 follows ${\bf
p}(ese)^2 \succ {\bf p}ese$, whence the node ${\bf p}(ese)^2$
exists.

The node $({\bf p}e,{\bf p}ese, {\bf p}(ese)^2)$ is an $SCC$-node
of  $\Gamma^3$ because all components of the node have common
right unit $e$. Let us notice that ${\bf p}(ese)^2  \ne  {\bf
p}ese$ and ${\bf p}e \ne {\bf p}ese$. We have $({\bf p}e, {\bf
p}ese) \succ ({\bf p}ese,{\bf p}(ese)^2)$. In the case ${\bf p}e =
{\bf p}(ese)^2$ we have contradiction with condition 1, in
opposite case we have contradiction with condition 3.

\subsection{Right local testability}

\begin{thm} $\label {4.1}$
 Let $S$ be transition  semigroup of
deterministic finite automaton  with state transition graph
$\Gamma$. Then $S$ is right locally testable iff

1. for any $SCC$-node
 ($\bf p, q$) from $\Gamma^2$ such that
 ${\bf p} \sim \bf q$ holds ${\bf p} = \bf q$.

 2. for any $SCC$-node
 ($\bf p, q$) $\in \Gamma^2$ and $s \in S$
 from  ${\bf p}s \succeq \bf q$
   follows ${\bf q}s \succeq \bf q$.
\end{thm}

Proof. Suppose semigroup $S$ is right locally testable.

Condition 1. Let  ($\bf p, q$) be an $SCC$-node  with distinct
components. Then for some idempotent $e \in S$ holds (${\bf p,
q})e= (\bf p, q$). If ${\bf p} \sim \bf q$ then for some $a, b \in
S$ holds ${\bf q}a = \bf p$ and  ${\bf p}b= \bf  q$, whence ${\bf
q}eae = \bf p$ and ${\bf p}ebe= \bf q$. So ${\bf q}eaebe = \bf q$
and ${\bf p}ebeae = \bf p$. Semigroup $S$ is right locally
testable and therefore the subsemigroup $eSe$ satisfies identity
$xyx=xy$ {\cite {GR}}. Consequently, ${\bf q = }{\bf q}eaebe= {\bf
q}eaebeae ={\bf p}ebeae = \bf p$.

Condition 2 follows from lemma \ref {3.3} because right locally
testable semigroup $S$ is locally idempotent.

Suppose now that both conditions of the theorem are valid. Let us
begin from the local idempotency of $S$.

If the identity $x^2=x$ is not valid in $eSe$ for some idempotent
$e$ then for some  node ${\bf v} \in \Gamma$
 and some element $a \in S$ we have ${\bf v}eae \ne {\bf v}eaeae$.
At least one of two considered nodes exists. In view of ${\bf v}e
\succeq {\bf v}eae \succeq {\bf v}eaeae$ the nodes ${\bf v}eae,
{\bf v}e$ exist. Let us denote
  ${\bf p = v}e, {\bf q = v}eae$. Therefore ($\bf p,
q$) is an $SCC$-node. Notice that ${\bf p}eae \succeq \bf q$.
Hence, by condition 2,
 ${\bf q}eae \succeq \bf q$. Now, by by condition 1, in view of
${\bf q} \succeq {\bf q}eae$, we have ${\bf q}eae = \bf q$. So
${\bf v}eae ={\bf v}eaeae$ in spite of our assumption.

Thus the transition semigroup $S$ is locally idempotent.

 If the identity $xyx=xy$ {\cite {GR}} is not valid  in $eSe$ then
for some node ${\bf v} \in \Gamma$, some idempotent $e$ and
elements $a, b \in S$ holds ${\bf v}eaebe \ne  {\bf v}eaebeae$. So
the node ${\bf v}eaebe$ exists. Let us denote ${\bf p = v}eaebe$.
$S$ is locally idempotent and therefore ${\bf p = v}eaebeaebe$.
Consequently, the node
 ${\bf q = v}eaebeae$ exists too. We have ${\bf p} \ne {\bf q}$.
  The node (${\bf v}eaebe, {\bf v}eaebeae)=(\bf p, q$) is an $SCC$-node
from $\Gamma^2$. It is clear that ${\bf p }={\bf v}eaebe \succeq
{\bf v}eaebeae = \bf q$. Then ${\bf q} = {\bf v}eaebeae \succeq
{\bf v}eaebeaebe = {\bf v}eaebe =\bf p$. So ${\bf p} \sim {\bf q}$
and ${\bf p} \ne {\bf q}$ in spite of the condition 1.

\subsection{Left local testability}
  \begin{lem} $\label {l1}$
 Let reduced $DFA$ {\bf A} with state transition graph $\Gamma$
 and transition semigroup $S$
be left locally testable. Suppose that for $SCC$-node ($\bf  p, q$)
of $\Gamma^2$ holds ${\bf p} \succeq \bf q$.

  Then for any $s \in S$
 holds ${\bf p}s \succeq \bf q$ iff ${\bf q}s \succeq \bf q$.
  \end{lem}

    Proof. Suppose {\bf A} is left locally testable.
 Then the transition semigroup $S$ of the automaton is
finite, aperiodic and for any idempotent $e \in S$ the
subsemigroup $eSe$ is idempotent {\cite{GR}}.

 For some $a, e=e^2 \in S$ holds ${\bf p}a = \bf q$,
  (${\bf p,q})e = (\bf p,q$).
  So we have ${\bf p}es={\bf p}s$ and ${\bf q}es = {\bf q}s$.

 If we assume that ${\bf p}s \succeq \bf q$, then for some $b$
from $S$ holds ${\bf p}sb= \bf q$, whence ${\bf p}esbe = \bf q$.
 In idempotent subsemigroup $eSe$ we have $esbe=(esbe)^2$.
 Therefore  ${\bf q}esbe = {\bf p}(esbe)^2 = {\bf p}esbe = \bf q$
and ${\bf q}es = {\bf q}s \succeq \bf q$.

  If we assume now that ${\bf q}s \succeq \bf q$, then for some $d \in S$
holds ${\bf q}sde = \bf q$. For some $a \in S$ holds ${\bf p}a  =
\bf q$ because of ${\bf p} \succeq \bf q$. So  ${\bf q}sde = {\bf
q}esde = \bf q$ and ${\bf p}eaesde = \bf q$. The subsemigroup
$eSe$ satisfies identity $xyx=yx$, therefore $eaesde = esdeaesde$.
So ${\bf q} = {\bf p}eaesde ={\bf p}esdeaesde$. Hence,  ${\bf p}es
={\bf p}s \succeq \bf q$.

\begin{lem} $\label {l2}$
 Let reduced $DFA$ {\bf A} with state transition graph $\Gamma$
be left locally testable.

 If the node ($\bf p, q, r$) is an
$SCC$-node of $\Gamma^3$,  (${\bf p, r}) \succeq (\bf q, r$) and
(${\bf p, q}) \succeq (\bf r, q$) in $\Gamma^2$, then  ${\bf r} =
\bf q$.
\end{lem}

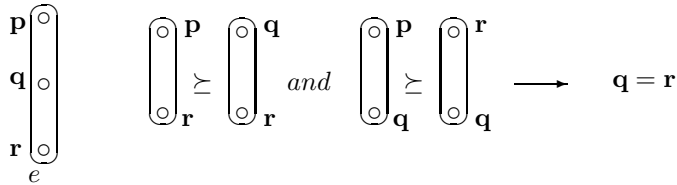
\begin{picture}(50,66)

\end{picture}
\begin{picture}(230,66)
\put(5,15){\circle{4}}   \put(-8,13){$\bf r$}
\put(5,40){\circle{4}}     \put(-8,40){$\bf q$}
\put(5,65){\circle{4}}    \put(-8,61){$\bf p$}
 \put(-1,3){$e$}
\put(5,40){\oval(10,60)}

\put(50,29){\circle{4}} \put(57,24){$\bf r$}
 \put(50,60){\circle{4}} \put(58,60){$\bf p$}
\put(50,45){\oval(10,40)}

\put(61,38){$\succeq$}

\put(80,29){\circle{4}} \put(88,24){$\bf r$}
 \put(80,60){\circle{4}} \put(88,60){$\bf q$}
\put(80,45){\oval(10,40)}

\put(97,38){$and$}

 \put(130,29){\circle{4}} \put(137,24){$\bf q$}
 \put(130,60){\circle{4}} \put(138,60){$\bf p$}
\put(130,45){\oval(10,40)}

\put(141,38){$\succeq$}

\put(160,29){\circle{4}} \put(168,24){$\bf q$}
 \put(160,60){\circle{4}} \put(168,60){$\bf r$}
\put(160,45){\oval(10,40)}

 \put(183,40){\vector(1,0){20}}

 \put(220,39){$\bf q = r$}

   \end{picture}

Proof. Suppose  {\bf A} is left locally testable.
 Then the transition semigroup $S$ of the automaton is
finite, aperiodic and for any idempotent $e \in S$ the
subsemigroup $eSe$ is idempotent {\cite {GR}}.

Let us consider the nodes $\bf  p, q, r$ from $\Gamma$ such that
the conditions of lemma are valid for them. From (${\bf p, r})
\succeq (\bf q, r$) and (${\bf p, q}) \succeq (\bf r, q$)  follows
(${\bf p, r})s = (\bf q, r$) and (${\bf p, q})t= (\bf r, q$)
 for some $s,t \in S$ and (${\bf p, q, r})e=(\bf p, q, r$),
  for some idempotents $e \in S$. We can take $s,t$ from $eSe$.
  Therefore

\centerline{$ese=s, ete=t, s^2 =s, t^2 =t$}

So ${\bf p}s = \bf q$, ${\bf r}s = {\bf r}$, ${\bf p}t = \bf r$,
${\bf q}t = \bf q$. Let us notice that
 ${\bf q}s={\bf p}s^2  = {\bf p}s = {\bf q}$. Analogously,
  ${\bf r}t={\bf r}$.

  We have  ${\bf p}sts ={\bf q}ts ={\bf q}s={\bf q}$.
  Then ${\bf p}ts = {\bf r}s = {\bf r}$. The identity $xyx=yx$ is
  valid in subsemigroup $eSe$, whence
 ${\bf q} = {\bf p}sts ={\bf p}ts ={\bf r}$.

\medskip
Let us formulate the necessary and sufficient conditions
for graph to be transition graph of DFA with
left  locally testable transition semigroup.

\begin{thm} $\label {l3}$
 Let $S$ be transition semigroup of a
deterministic finite automaton  with state transition graph $\Gamma$.

Then   $S$ is left locally testable iff

1. $S$ is locally idempotent,

2. for any $SCC$-node ($\bf  p, q$) of $\Gamma^2$
 such that ${\bf p} \succeq \bf q$ and for any $s \in S$
 we have ${\bf p}s \succeq \bf q$ iff ${\bf q}s \succeq \bf q$ and

3. If for arbitrary nodes ${\bf  p, q, r} \in \Gamma$ the node
($\bf p, q, r$) is $SCC$-node of $\Gamma^3$, (${\bf p, r}) \succeq
(\bf q, r$) and (${\bf p, q}) \succeq (\bf r, q$) in $\Gamma^2$,
then  ${\bf r} = \bf q$.
\end{thm}

Proof. Suppose semigroup $S$ is left locally testable. Then $S$ is
locally idempotent {\cite {GR}}. Second and third conditions of
our theorem follow from lemmas $\ref {l1}$ and $\ref {l2}$,
correspondingly.

Suppose now that the conditions of the theorem are valid but for
an arbitrary node $\bf p$, an arbitrary idempotent $e \in S$ and
two elements $s, t \in eSe$ holds ${\bf p}sts \ne {\bf p}ts$.
 By condition 1,

\centerline{$s^2=s$, $t^2=t$, $tsts=ts$, $stst=st$, $tssts=ts$}

 At least one of two nodes ${\bf p}sts = \bf q$ and
 ${\bf p}ts = \bf r$ exists. Therefore ${\bf p}e$ exists too.
 We have (${\bf  p}e, {\bf p}ts)sts =({\bf  p}sts, {\bf p}ts$).
 Therefore the existence of the node  ${\bf p}ts = \bf r$
implies by condition 2 the existence of the node  ${\bf p}sts =
\bf q$. Analogously, from (${\bf  p}e, {\bf p}sts)ts =({\bf p}ts,
{\bf p}sts$) and existence of the node  ${\bf p}sts = \bf q$
follows by condition 2 the existence of the node  ${\bf p}ts = \bf
r$.

The node (${\bf  p}e, {\bf q, r}$) is an $SCC$-node because all
his components have common right unit $e$. We have (${\bf p,
r})sts = ({\bf p}sts, {\bf p}tssts) = ({\bf q}, {\bf p}ts) =(\bf
q, r$). Analogously, $({\bf p, q})ts = ({\bf p}ts, {\bf p}ststs) =
({\bf r}, {\bf p}sts) =(\bf r, q$). Thus,

 \centerline{$({\bf p}e, {\bf r}) \succ (\bf q, r)$,
  $({\bf p}e, {\bf q}) \succ (\bf r, q)$}

Now by the third condition of the theorem, ${\bf r} = \bf q$.
Therefore ${\bf p}sts = {\bf p}ts$. The node $\bf p$ is an
arbitrary node, whence $sts=ts$ for every two elements $s,t \in
eSe$. Consequently, the subsemigroup $eSe$ satisfies identity
$xyx=yx$. Thus the semigroup $S$ is left locally testable.

  \section{Semigroups}

\begin{lem} {\label {s1}}
Let $S$ be a finite locally idempotent semigroup.
The following two conditions are equivalent in $S$:

{\bf a)} $S$ satisfies locally the identity  $xyx=xy$ ($S$ is
right locally testable).

{\bf b)} No two distinct idempotents $e$, $i$ from $S$ such
that  $ie=e,ei=i$ have a common  right unit in $S$. That is, there is no
idempotent $f \in S$ such that $ e =ef$ and $i =if$.
\end{lem}

    Proof. Suppose the identity $xy=xyx$ is valid in
subsemigroup $uSu$ for any idempotent $u$
 and for some idempotents $e$, $i$ in $S$ we have $ie=e$, $ei=i$.
 Suppose $f$ is a common right unit of $e$, $i$.
 The identity $xyx=xy$ in $fSf$ and equality $ei=i$
 imply $i=ei=efefif=efefifef=eie=e$.
 Thus the idempotents $e$, $i$ are not distinct.

   Suppose now that $uSu$ does not satisfy the identity $xyx=xy$
 for some idempotent $u$. Notice that $uSu$ is an  idempotent
semigroup.
 So for some $a$, $b$ of $S$, $uaubuau \neq uaubu$.
For two distinct idempotents $i=uaubuau$ and $e=uaubu$
with common right unit $u$ we have
 $ie=uaubuauuaubu=uaubuaubu=uaubu=e$ and
 $ei=uaubuaubuau=uaubua=i$.

So  two distinct idempotents $e$, $i$ from $S$ such
that  $ie=e, ei=i$ have a common  right unit  $u$ in $S$.

The following lemma is proved analogously:
\begin{lem} $\label {s2}$
Let $S$ be a finite locally idempotent semigroup.
The following two conditions are equivalent in $S$:

{\bf a)} $S$ satisfies locally the identity  $xyx=yx$ ($S$ is left
locally testable).

{\bf b)} No two distinct idempotents $e$, $i$ from $S$ such
that  $ie=i,ei=e$ have a common left unit in $S$. That is, there is no
idempotent $f \in S$ such that $ e =fe$ and $i =fi$.
\end{lem}

Recall that a semigroup $A$ is a right [left] zero semigroup if
$A$ satisfies the identity $xy = y [xy = x]$.
     A right [left] locally testable semigroup is
 locally idempotent  \cite {GR}.
 Then from the last two lemmas follows

\begin{thm} $\label {s3}$ A finite semigroup $S$
 is right [left] locally testable iff $S$ is locally
idempotent and no two distinct idempotents $e$, $i$ from
right [left] zero subsemigroup
 have a common right [left] unit in $S$.
\end{thm}

 \section{An algorithm for semigroup}

The following proposition is useful for the algorithm.

\begin{lem} $\label {s4}$  {\cite {Ta}}
Let $E$ be the set of idempotents of a semigroup $S$ of size
$n$ represented as an ordered list.
Then there exists an algorithm of order $n^2$ that
reorders the list so that the maximal left [right] zero subsemigroups of $S$
appear consecutively in the list.
\end{lem}

   1.{\it Testing whether a finite semigroup $S$ is right [left] locally
testable.}
 \medskip

 Suppose $|S| = k$. We begin by finding the set of idempotents $E$.
 This is a linear time algorithm.
Then let us verify local idempotency. For every $e \in E$ and
every $s \in S$ let us check condition $ese = (ese)^2$.
If the condition does not hold for some pair, the semigroup
is not locally idempotent and therefore not right locally
testable (theorem $\ref {s3}$). This takes  $O(k^2)$  steps.

   Now we reorder $E$ according to lemma {\ref {s4}} in a
 chain such that the subsemigroups of right [left] zeroes form
 intervals in this chain. We note the bounds of these
 intervals. We find for each element $e$ of $E$ the first element $i$ in
 the chain such that $e$ is a right [left] unit for $i$. Then we find in the chain
 the next element $j$ with the same unit $e$. If $i$ and $j$ belong
 to the same subsemigroup of right [left] zeroes we conclude that $S$
 is not right [left] testable (Lemma  {\ref {s1}}) and stop the process.
 If they are in different right [left] zero semigroups,
 we replace $i$ by $j$ and continue the process of finding a new $j$. This
 takes  $O(k^2)$  steps.

Finding the maximal subsemigroup of right [left] zeroes containing a given
idempotent needs $k$ steps. So for to reorder $E$ we need at most $k^2$ steps.
The time and the space complexity of the algorithm is $O(k^2)$.

\section{Graph  algorithms}
Let $n$ be the sum of the nodes and edges of $\Gamma$. The
first-depth search (\cite {AH}, \cite {K91} or \cite {TL}) will be
used for SCC search, for reachability table for triples and for
checking condition 2 of theorems \ref{3.8} and \ref{4.1}.

\medskip
{\it Table of reachability for triples}

Suppose $SCC$ of $\Gamma$,
 $\Gamma^3$ and the table of reachability  are known.
For every $SCC$-node {\bf q} of the graph $\Gamma$ let us
form by help of the first-depth search on $\Gamma^2$
 the following relation $L$ [I] on $\Gamma$:
 ${\bf p} L \bf r$ if ($\bf p, q) \succeq (\bf r, q$)
[${\bf p} I \bf r$ if ($\bf p, q) \succeq (\bf q, r$)]. For every
node ($\bf p, q)$ we form set of nodes $\bf r$ such that
${\bf p} L \bf r$ [${\bf p} I \bf r$]. We use an auxiliary array for this aim:
for every node ($\bf p, q)$ and for every node $\bf s$,
we form set of pointers to nearest successors ($\bf t, s$)
[($\bf s, t$)] of ($\bf p, q$). 

If ($\bf p, q, r$) is an $SCC$-node with distinct components
 and ${\bf p} L \bf r$ [${\bf p} I \bf r$]
then we add the triple
($\bf p, q, r$) to the set $Left$ [$LocId$].
($O(n^3)$ time and space complexity).

\subsection{ Graph of automaton with locally idempotent transition semigroup}
The algorithm is based on the theorem {\ref {3.8}}.
   Let us recognize the reachability on the graph $\Gamma$
 and form the table of reachability for all pairs
of $\Gamma$. The time required for this step is $O(|\Gamma|^2)$.

 We find graph $\Gamma^2$ and all $SCC$
 of the graph ($O(n^2)$ time complexity).
If the nodes ($\bf p, q$) and  ($\bf q, p$) belong to common $SCC$
then the transition semigroup is not locally idempotent (condition
1).

For check  the condition 2 of the theorem let us add to the graph
$\Gamma^2$ new node ($\bf 0, 0$) with edges from this node to
every $SCC$-node ($\bf p, q$) from $\Gamma^2$ such that ${\bf p}
\succeq \bf q$. Let us consider first-depth search from the node
($\bf 0,0$) (the unique starting point of any path).

  Let us fix the node $\bf q$ after going through the edge (${\bf 0, 0)} \to
 (\bf p, q$).
  We do not visit edges ($\bf r, s) \to (\bf r, s$)$\sigma$
 such that
${\bf r}\sigma \not\succeq \bf s$. In the case that for the node
($\bf r, s$) from two conditions ${\bf r}\sigma \succeq \bf q$ and
${\bf s}\sigma \succeq \bf q$ only the first is valid the
condition 2 does not hold, the transition semigroup is not locally
idempotent and the algorithm stops.

 Let us find graph $\Gamma^3$,
    all $SCC$ of the graph $\Gamma^3$
 and mark all $SCC$-nodes with three distinct components
 such that the first component is ancestor of two others.
($O(n^3)$ time complexity).

 Let us go to the condition 3 of the theorem {\ref {3.8}}.
 We form a table of triples $LocId$
 (see algorithm for table of reachability above).
 If some $SCC$-node ($\bf p, q, r$) from $\Gamma^3$
 with distinct components belongs to $LocId$ then
 the condition 3 does not hold and the semigroup is not locally
idempotent.

The whole time and space complexity of the algorithm is  $O(n^3)$.

\subsection{Right local testability of  DFA}
The algorithm is based on the theorem {\ref {4.1}}. Let us form a
table of reachability of the graph $\Gamma$, find all $SCC$ of
$\Gamma$, $\Gamma^2$ and all $SCC$-nodes of $\Gamma^2$. ($O(n^2)$
time complexity).

 Let us verify the condition 1 of the theorem.
  For every $SCC$-node ($\bf p, q$) (${\bf p} \neq \bf q$) from $\Gamma^2$
let us check the condition ${\bf p} \sim \bf q$. If the condition
holds the automaton is not right locally testable. ($O(n^2)$ time
complexity).

 For check  the condition 2 of the theorem
  let us add to the graph
$\Gamma^2$ new node ($\bf 0, 0$) with edges from this node to
every $SCC$-node ($\bf p, q$) from $\Gamma^2$ such that ${\bf p}
\succeq \bf q$. Let us consider first-depth search from the node
($\bf 0,0$) (the unique begin of any path).

  Let us fix the node $\bf q$ after going through the edge (${\bf 0, 0)} \to
 (\bf p, q$).
  We do not visit edges ($\bf r, s) \to (\bf r, s$)$\sigma$
 such that
${\bf r}\sigma \not\succeq \bf s$. In the case that for the node
($\bf r, s$) from two conditions ${\bf r}\sigma \succeq \bf q$ and
${\bf s}\sigma \succeq \bf q$ only the first is valid the
algorithm stops and the condition 2 does not hold. The automaton
is not right locally testable in this case.
 ($O(n^2)$ time complexity).

The whole time and space complexity of the algorithm is  $O(n^2)$.

\subsection{Left local testability of  DFA}
The algorithm is based on the theorem {\ref {l3}}. Let us form a
table of reachability on the graph $\Gamma$ and find all $SCC$ of
$\Gamma$.
 Let us find $\Gamma^2$ and all $SCC$ of $\Gamma^2$.
($O(n^2)$ time complexity).

Let us check the local idempotency
 ($O(n^3)$ time complexity).

 For check  the condition 2 of the theorem
  let us add to the graph
$\Gamma^2$ new node ($\bf 0, 0$) with edges from this node to
every $SCC$-node ($\bf p, q$) from $\Gamma^2$ such that ${\bf p}
\succeq \bf q$. Let us consider first-depth search from the node
($\bf 0,0$).

  We do not visit edges ($\bf r, s) \to (\bf r, s$)$\sigma$
 such that
${\bf r}\sigma \not\succeq \bf s$ and ${\bf s}\sigma \not\succeq
\bf s$. In the case that for the node ($\bf r, s$) from two
conditions ${\bf r}\sigma \succeq \bf s$ and ${\bf s}\sigma
\succeq \bf s$ only one is valid the algorithm stops and the
condition 2 does not hold.

Condition 3 of the theorem $\ref {l3}$.
 Let us find $\Gamma^3$ and all $SCC$-nodes of $\Gamma^3$
 ($O(n^3)$ time complexity).

 Let us recognize the relation $\succ$ on the graph $\Gamma^2$
 and find set $Left$ of triples  $\bf p, q, r$ such that
 (${\bf p, q}) \succ (\bf r, q$) (see algorithm for table of reachability above).

  If for some $SCC$-node ($ \bf p, u, v$) of $\Gamma^3$ both triples
    ($\bf p, u, v$) and ($\bf p, v, u$) belong to the set
     then the condition 3 does not hold,
  the automaton is not left locally testable and the algorithm stops.

The whole time and space complexity of the algorithm is $O(n^3)$.

 \end{document}